\documentclass[a4paper]{article}

\usepackage{amsmath}
\usepackage{amssymb}
\usepackage{latexsym}
\usepackage{graphicx}
\usepackage{pb-diagram}

\usepackage[all]{xy}

\usepackage{natbib} 
\usepackage{makeidx}
\usepackage{soul}
\usepackage[usenames,dvipsnames]{color}
\usepackage{pdfcolmk} 	
\usepackage[colorlinks,urlcolor=MidnightBlue,citecolor=Sepia,linkcolor=Maroon,pdfstartview=FitH,pdfpagemode=UseNone]{hyperref}

\newcounter{ncomm}

\makeindex

\begin{document}

\title{Causality in concurrent systems}

\author{Silvia Crafa \\
       {\small Dipartimento di Matematica}  \\
       {\small Universit\`a di Padova, Italy}  
    \and
       Federica Russo\\ {\small Center Leo Apostel} \\ {\small Vrije Universiteit Brussel} }
\date{}
\maketitle

\begin{abstract}

Concurrent systems identify systems, either software, hardware or even
biological systems, that are characterized by sets of independent
actions that can be executed in any order or simultaneously. Computer
scientists resort to a causal terminology to describe and analyse the
relations between the actions in these systems. However, a thorough
discussion about the meaning of causality in such a context has not
been developed yet. This paper aims to fill the gap. First, the paper
analyses the notion of causation in concurrent systems and attempts to
build bridges with the existing philosophical literature, highlighting
similarities and divergences between them. Second, the paper analyses
the use of counterfactual reasoning in ex-post analysis in concurrent
systems (i.e.~execution trace analysis).

\end{abstract}

\section{Introduction}

In the terminology of computer science, Concurrent Systems identify
systems, either software, hardware or even biological systems, where
sets of activities run in parallel with possible occasional
interactions. A simple example of concurrent system is the Internet,
which can be thought of as a set of computers, each one computing its
independent activity, that often communicate to exchange some
information. A further example is the railway system of a country,
where many trains travel sharing tracks in an ordered way so that two
trains can move at the same time along different tracks, whereas a
single track (e.g, a platform in a train station) can only be used by
a single train at a time. Furthermore, the large number of activities
carried on by a single human cell form a biological concurrent system,
that actually shares a number of similarities with the Internet.

Compared to sequential systems, where a single action is executed at
a time according to a sequential algorithm, concurrent systems raise
new complex issues dealing with the ordering of action executions. 
In particular, concurrent systems are characterized by sets of 
independent actions that can be executed in any order or
simultaneously. As a consequence computer scientists resorted to the
causal terminology to describe and analyse the relations between the
system actions. However, a thorough discussion about the meaning of
causality in such a context has not been developed yet.  

In this paper, therefore, we ask precisely what causality means and
how causal reasoning works in concurrent systems. As we will show in
the foregoing discussion, causality here means quite generally
\emph{dependence}, whether temporal, spacial, or even causal dependence. 
We will also see that counterfactual reasoning is indeed one of the
causal tools used in concurrent systems. 

The paper is organised as follows. In section \S \ref{sec:ConcSysts}
we provide an introduction to concurrent systems, in particular we
explain their formal language and operational models through simple
examples. In \S \ref{causality} we explain how causal relations are
formally defined in concurrent systems and what causality means
in this context. We also discuss this specific meaning with respect to
more traditional debates in the philosophy of causality, for instance
about production and mechanisms, independence, and causation by
omission. In \S \ref{analysis} we discuss analysis techniques in
concurrent systems, showing that
counterfactual reasoning is a useful tool in this context. We provide
examples of counterfactual validation and refutation using the theory
of Nicholas Rescher.

\section{Concurrent Systems: a Crash Course}
\label{sec:ConcSysts}

Concurrent Systems identify systems where sets of activities run in
parallel with possible occasional interactions. Moreover, new
activities can be dynamically added to the system during
its evolution, and terminated or aborted activities can be removed as
well, possibly in unexpected ways.  

In computer science,  a simple example of concurrent system is a
network like Internet: it can be thought of as a set of computers,
each one computing its independent activity, that often communicate to
exchange some information and that unpredictably disconnect.
The activities of (mobile) phones in a town, the railway system of a
country, or else the large number of activities carried on by a single
human cell are also concurrent systems that share similarities
with the Internet. To illustrate, we discuss some (simple) examples. 

\paragraph{Railway system.} Consider the simple railway system
depicted below, where we assume that each pair of stations is
connected by a single track. 
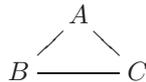
\begin{center}
\mbox{
\xymatrix@C=3mm@R=3mm{
    & A \ar@{-}[dl]\ar@{-}[dr] & \\
B \ar@{-}[rr]  &    & C
}
}
\end{center}
Then consider Train1 that leaves form $A$, reaches $B$ and then $C$,
and Train2 that leaves form $A$, reaches $C$ and then $B$. The
two trains can move concurrently (possibly at different speed) 
to their first stop, while the transit on the track between $B$ and
$C$ must be regulated so that they do not collide. 
In other words, the presence of Train1 at 
$C$ \emph{depends} on the previous presence of Train1 at $B$.
Moreover, the usage of track $AB$ by Train1 and
track $AC$ by Train2 are \emph{concurrent} activities, that can take
place in any order or at the same time as well. 
Finally, we say that the usage of track $BC$ by Train1 is \emph{in
  conflict} with the usage of the same track by Train2. Hence the
track BC can only be traveled across either by first Train1 and then
by Train2 or vice-versa, and the two possibilities are equally valid.
That is, in absence of a fixed train scheduling,
the choice between these two possibilities
is called \emph{nondeterministic}. 
A nondeterministic choice between a set of possible alternatives is a
choice that can be solved in different ways during different system
executions. For instance, in one execution of the railway system first
Train1 travels along BC and then Train2, but in a different execution
Train2 is the first one travelling across BC.
Note that concurrent systems are generally nondeterministic, meaning
that different executions of the same system may lead to different
behaviours according to different runtime choices between
nondeterministic alternatives, each of them equally valid even if
observationally different.


\paragraph{Computer network.} As a further example, consider the following computer network, where we
assume that computers exchange messages:
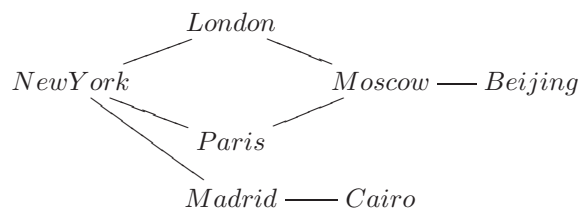
\begin{center}
\mbox{
\xymatrix@C=5mm@R=3mm{
         &  \ar@{-}[dl] London \ar@{-}[dr] &   &     \\
New York \ar@{-}[ddr]&  &  Moscow \ar@{-}[r] & Beijing \\
             &  \ar@{-}[ul] Paris \ar@{-}[ur]&   &  \\
           & Madrid \ar@{-}[r] & Cairo &
}}
\end{center}
Messages can travel concurrently along different links and
sequentially on the same link. For instance, a message generated in
New York can be sent to Cairo and at the same time a different message
can be sent to Beijing travelling either through London or Paris.
In this case the reception of the message in
Beijing ``causally'' depends on the fact that it has passed through
London or Paris (we shall discuss later what ``causally'' mean).  
Notice that such ``or-causality'' is not an issue since it might
be ok to abstract from---in the sense of not knowing---the precise path 
that the message has followed, as long as we are interested in relaxed
properties such as ``Europe might have sniffed the message in
transit'' or ``Egypt has certainly have not sniffed the message''.

\paragraph{Phone calls and text messages.}
As a final example, we distinguish from synchronous and asynchronous
communication. On the one hand, the way two people talk through a phone call is a
\emph{synchronous} communication, that is speaking and listening happens at
the same time. On the other hand, texts sent with mobile phones are
\emph{asynchronous} communications since the writer does not assume that the
reading is immediate, hence he can proceed with its activity without
waiting for the receiver to read the text. In the 
asynchronous scenario there are less ``causal'' dependencies than in
the synchronous one. To illustrate, consider Alice that wants to
communicate a three paragraph length letter to Bob. In a phone call
Bob will not listen the final paragraph before he has listened the
first one. However, if Alice writes the letter to Bob by means of a
long text, it might happen that the phone splits the long text into
three texts of fixed length, more or less corresponding to the three
paragraphs. Even if the three texts are sent by Alice in the right
order, there are no constraints on the order in which the three
paragraphs are received by Bob. Indeed, it is not a problem for Bob 
to first receive the final paragraph, since he eventually
receives the remaining texts and they are properly recomposed. Hence,
in an asynchronous system the order of reception does not
\emph{depend} on the order of sending.

\bigskip \noindent
In computer science, the complexity of (concurrent) systems is 
addressed in a standard way: first the system is specified using the
syntax of a precise formal language; then, the system behaviour is 
formally described by means of a semantic operational model; finally, analysis techniques and formal reasoning are developed (and
automated) to prove properties of such a system on top of its
operational model. As far as concurrent systems are concerned, 
a number of formalisms have been proposed; they can be
summarized as follows, according to three different levels of
abstraction described above:
\begin{description}
\item[Formal languages] to specify/to program a concurrent
  system. Examples of these languages are the Java programming
  language, that is commonly used to write concurrent software, 
  and a number of domain
  specific languages targeted to the description of specific
  concurrent systems such as security protocols, concurrent hardware,
  system biology. Besides a rich number of ``concrete concurrent
  languages'', the research on concurrent theory has distilled few
  formal languages made of a minimal number of
  primitives/connectives that capture the essential features of 
  concurrent systems \citep{CCS,pi}. 
   Such languages are also called \emph{process
    algebras}
  to highlight the mathematical treatment of connectives like
  sequential composition, parallel composition, interaction,
  nondeterministic choice. 
\item[Operational models] describing all the possible behaviours of a
  given concurrent system (written in a formal language). 
  Many different formalisms have been proposed to define such models,
  corresponding to different trade-offs between the expressivity of the
  model and its simplicity/abstra\-ction power.
  Standard models for concurrent systems are based on the
  \emph{interleaving} approach: they are based on the idea that 
  only a single action can be executed at a time, according to what
  actually happens in a single-processor computer. In these models
  two concurrent actions $A$ and $B$ are modeled as the
  nondeterministic choice between executing first $A$ and then $B$ or
  viceversa, i.e. the arbitrary interleaving of $A$ and $B$ rather
  than their simultaneous occurrence. Reducing the
  notion of concurrency to those of nondeterminism and sequentiality
  allows us to build models that are easier to deal with, and has provided the
  basis for the development of very rich and elegant theories of
  concurrency (see for instance \citet{ReactiveSyst}). 
  However, there are properties of concurrent systems
  that cannot be specified without a clear distinction between
  concurrency and nondeterminism. Then a number of
  non-interleaving models, a.k.a.~\emph{true-concurrent models} and
  sometimes \emph{causal models}, have
  been developed by taking the notion of concurrency---or the
  complementary notion of causality---as fundamental \citep{WN95}. 
  We discuss below the operational models of the previous
  examples. For the time being, it will suffice to say that an interleaving model of the railway
  system described in the previous section cannot model the case where
  Train1 is travelling across $AB$ and at the same time Train2 is
  travelling across $AC$; in such a model the two actions can be
  executed in any order but not at the same time. Instead, a
  true-concurrent model can express all the possible behaviours:  
  Train1 travels before Train2, Train2 travels before Train1, and also
  Train1 and Train2 travel at the same time.

\item[Analysis techniques.] Working with a model
  that explicitly talks about (absence of) causal dependencies
  between the actions of the system allows  formal reasoning about such
  dependencies. A rich toolbox of formal methods have been developed
  to specify and automatically verify (causal) properties of models of
  concurrent systems. Model checking tools, diagnosis methods
  examining the causal history 
  of an error occurrence, behavioural abstractions based on the
  observable degree of concurrency, and specification temporal logics 
  are examples of such formal analysis techniques.  
\end{description}

\section{Causal Concepts in Concurrent Systems}
\label{causality}

In this section we explain how `causal relations' are defined in concurrent systems and we discuss the meaning of causality in this context. It will emerge that concurrent systems are quite peculiar with respect to many other areas investigated in the philosophy of causality. In fact, causal relations are defined and decided by the programmer, rather than discovered or established as customarily done by biologists, social scientists, or physicists. Causal talk may also appear `loose', or even unnecessary, as causality means, in this context, just dependence but not production.

\subsection{Causal Relations in Concurrent Systems}

We start by looking at how the term ``causality'' is customarily used
in the semantics of concurrent systems. 
First, since in computer science everything
is discrete, we call \emph{event} the occurrence of an action of the
systems, i.e.~a step of computation. Hence, in sequential programs,
where there is a single flow of control at any time, causality is
interpreted as space-time sequentiality of events.  
The case of concurrent systems is more complex:
as anticipated above, for these systems there are both
interleaving and true-concurrent models. We are now interested in the
second class of models, that can express the difference between 
two actions that might simultaneously happen and two actions that can
just be executed in any order. True concurrent models include, 
between others, Petri nets, event structures, generalized
labelled transition systems and causal trees. Each of them comes 
equipped with its specific formal analysis techniques, but their
expressivity is roughly the same \citep{WN95}, 
hence we briefly review here Prime Event Structures, which are in a
sense canonical ad pretty accessible. 


A Labelled Prime Event Structure ${\cal E}$ is a tuple 
$\langle E,\leq,\#,\lambda\rangle$ where 
\begin{itemize}
\item $E$ is a set of events and $\lambda$ is a function that
  associates to each event the action whose occurrence that event
  stands for;  
\item $\leq$ is a partial order representing the
  \emph{causal relation} between events. Let be $e_1,e_2\in E$, then
  we write $e_1\leq e_2$ to state that $e_1$ is a cause of $e_2$.  
  By definition of partial order, the causal relation is
  reflexive, i.e. $e\leq e$, transitive, i.e. if
  $e_1\leq e_2$ and $e_2\leq e_3$ then $e_1\leq e_3$ and
  antisymmetric, i.e. if $e_1\leq e_2$ and $e_2\leq e_1$ then
  $e_1=e_2$. When the set of
  events $E$ is infinite (infinite computation), we assume that each
  event has only finitely many causes, i.e.~causal histories are
  finite. Later, we shall discuss in  detail what ``causality'' means in this context.
\item $\#$ is an irreflexive and symmetric relation called
  \emph{conflict}.  Let be $e_1,e_2\in E$, then we write
  $e_1\# e_2$ to express two alternative behaviours: either the system
  executes $e_1$ or it executes $e_2$, the choice between the two
  alternative behaviours is purely nondeterministic. As an example,
  remember the single track railway system where Train1 travelling
  across the track $BC$ is in conflict with Train2 travelling across
  the same track (we give below the complete event structure model of
  the railway system).  There is an additional axiom stating that the
  conflict is hereditary, that is if $e_1$ is a cause of $e_2$ and
  $e_1$ is in conflict with $e_3$, then $e_2$ must also be in conflict
  with $e_3$. 
\end{itemize}

Let us first illustrate these true-concurrent models by
defining the prime event structures associated to the 
examples given in Section~\ref{sec:ConcSysts}.
The discussion about the meaning of the causal relation introduced by
event structures is postponed to the next subsection.

\paragraph{Computer network.} We begin by illustrating the event structure associated to the transmission of
two messages from New York to both Beijing and Cairo in the
computer network described above.

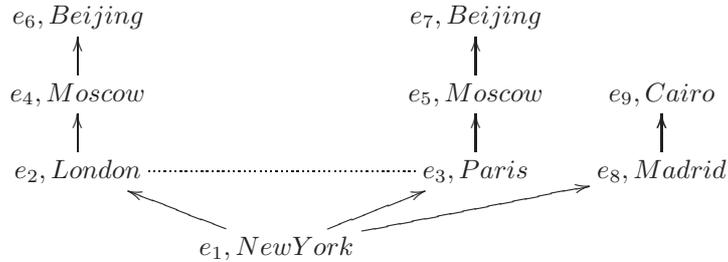
\begin{center}
\mbox{
\xymatrix@C=5mm@R=5mm{ 
\ar@{<-}[d] e_6, Beijing &  & \ar@{<-}[d] e_7, Beijing   & \\
  \ar@{<-}[d] e_4, Moscow &  & \ar@{<-}[d] e_5, Moscow   & \ar@{<-}[d]e_9, Cairo    \\
 \ar@{<-}[dr] e_2, London \ar@{.}[rr]&  & \ar@{<-}[dl] e_3, Paris &
e_8, Madrid \ar@{->}[u]\\
       & e_1, New York \ar@{->}[urr]&    \\
}}
\end{center}

In the figure ``causal relations'' proceed upwards following the
arrows; ``conflict'' is depicted by dotted lines. The event
structure contains 9 events whose labels are specified besides
events. The event $e_1$ (corresponding to message creation in New
York) is a cause of $e_2,e_3,e_8$, since in any behaviour of the
system the event $e_2$ (similarly for $e_3$ and $e_8$) cannot occur
without a previous occurrence of $e_1$, namely it cannot be the case
that a message is in London, Paris, Madrid if it was not in New York
before. The events $e_2$ and $e_3$ are in conflict, accordingly, the
path from New York to Beijing nondeterministically crosses either
London or Paris. On the other hand, events $e_4$ and $e_5$ have the
same label, both standing for the presence of the message in
Moscow. However, they are kept different since they have a different
causal history: $e_4$ stands for the arrival of the message in Moscow
travelling across London, while $e_5$ stands for the arrival of the
message along the alternative path. This example illustrate how
disjoint causality is represented in event structures by duplicating
events (with the same label but different causal histories). We shall get back to disjoint structure later in section \ref{subsec:meaning}.

Notice  that
there is a conflict also between $e_4$ and $e_5$ (as well as between
$e_2$ and $e_5$, between $e_3$ and $e_4$, and similarly for $e_6$ and
$e_7$), however this is a conflict inherited from the conflict between $e_2$
and $e_3$, but we do not depict inherited conflicts in order to keep
the picture more readable.

Finally, the events $e_2$ and $e_8$ are
not causally related nor in conflict;  they are then said to be
\emph{concurrent}. Indeed, in our system it is possible that a 
message is in Madrid, travelling to Cairo, while the other message is
in London travelling to Beijing. Notice that there are many other
pairs of concurrent events, such as ($e_6$,$e_8$), ($e_9$,$e_2$),
($e_5$ and $e_9$), etc. Recall, concurrent events are independent:
they can occur in any order and also simultaneously. 

\paragraph{Phone calls and text messages.}
Consider  the  two event
structures corresponding to the system made of Alice and Bob, where
Alice first performs the action A1, then talks to Bob and continues
with the action A2, while Bob first performs the action B1, then
listens to Alice and continues with the action B2. 

 \begin{center}
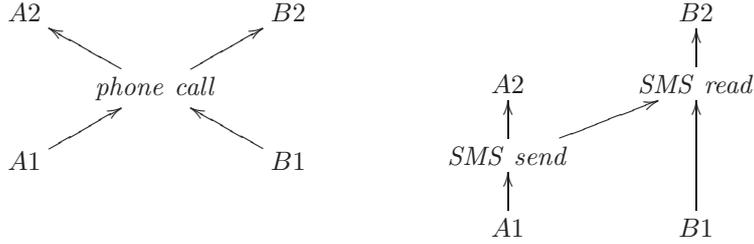
\begin{tabular}{ll}
\mbox{
\xymatrix@C=5mm@R=5mm{
  A2 &  & B2 \\
     & \mathit{phone\ call}\ar@{->}[ur]\ar@{->}[ul] &  \\
  A1\ar@{->}[ur] &  & B1\ar@{->}[ul]
}}
&
\hspace{1cm}
\mbox{
\xymatrix@C=7mm@R=5mm{
   &   B2 \\
  A2  &  \mathit{SMS\ read} \ar@{->}[u]\\
  \mathit{SMS\ send}\ar@{->}[ur]\ar@{->}[u] &   \\
  A1\ar@{->}[u]   & B1\ar@{->}[uu]
}}
\end{tabular}
\end{center}

The event structure in the left represents the synchronous
communication between Alice and Bob, while the event structure in the
right represents the asynchronous send of a text. Accordingly, only in
the asynchronous case Alice can perform the action A2 without even
waiting for Bob's completion of action B1. Notice that in both
cases the occurrence of Bob's B2 action depends on the previous
occurrence of Alice's A1 action, i.e.~A1 is in the causal history of
B2.

\paragraph{Railway system.}
The case of the single-track railway system is more complex. We can
represent its causal model with the following event structure:

 \begin{center}
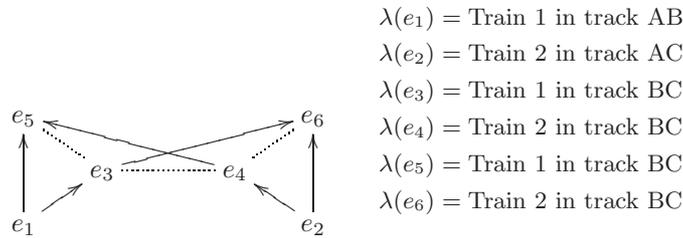
\begin{tabular}{ll}
\mbox{
\xymatrix@C=5mm@R=3mm{
  \ar@{.}[dr] \ar@{<-}[drrr]e_5  & & & & \ar@{<-}[dlll] e_6  \ar@{.}[dl]  \\
           & e_3 \ar@{.}[rr]& & e_4 &        \\
       e_1  \ar@{->}[uu] \ar@{->}[ur]  & &   &      & e_2 \ar@{->}[uu] \ar@{->}[ul]    \\
}}&
{\small
$
\begin{array}{l}
\lambda(e_1) = \mbox{Train 1 in track AB} \\[1mm] 
\lambda(e_2) = \mbox{Train 2 in track AC} \\[1mm]
\lambda(e_3) = \mbox{Train 1 in track BC} \\[1mm]
\lambda(e_4) = \mbox{Train 2 in track BC} \\[1mm]
\lambda(e_5) = \mbox{Train 1 in track BC} \\[1mm]
\lambda(e_6) = \mbox{Train 2 in track BC} 
\end{array}
$
}
\end{tabular}
\end{center}

This model illustrates which events can occur in parallel (or in any
order) and which event must occur before another. For instance
$e_1$ and $e_2$ are concurrent, that is Train 1 and Train 2 can freely
travel across their first track, either at the same time, or in any
order. However, Train 2 cannot travel across BC before
travelling across AC, represented by the dependency of, e.g., $e_6$ on
$e_2$. Moreover, both events $e_3$ and $e_5$ represent Train 1 in
track BC. They both depend on $e_1$, i.e. the passage of Train 1 in
track AB, but $e_5$ also depend on $e_4$, i.e. the passage of Train 2
in track BC. In other words, $e_3$ represent Train 1 that first
travels across track BC, while $e_5$ represent Train 1 that travels
across BC after Train 2. The conflict between $e_5$ and $e_3$ models
the fact that Train 1 either travels across BC first or second. 
On the other hand the conflict between $e_3$ and $e_4$ models the
nondeterminism of the first Train starting its travel across the track
BC.

\bigskip
\noindent It should be clear by now that the complexity of causal models rapidly grows:
even the behaviour of simple concurrent systems corresponds to big
event structures due to all the nondeterministic interleaving of
concurrent actions and the duplication of events with disjoint causes.
Moreover, concurrent systems usually have infinite behaviour (e.g. a
railway system does not (should not) stop) which can be finitely
represented by a finite set of states the system can enter infinitely
many times. Despite such complexity, formal methods have been developed
to automatically derive a true-concurrent model associated to a
given concurrent program/system specified in a handy concurrent
language, and further improvements are subject of current research.

\subsection{The meaning of causality in concurrent systems}
\label{subsec:meaning}

\paragraph{Causality as a primitive relation.}
The main point to notice is that in Prime Event Structures 
causality is taken to be a \emph{primitive relation}: given an event
structure ${\cal E}$, an event $A\in E$ is
causally dependent on an event $B\in E$ if and only if it has been
so defined, i.e. if and only if $(B,A)\in\, \leq$.
Indeed, these models are not intended to be used for causal discovery:
instead of asking whether two events are related by a causal relation
or not, which might be difficult or controversial, we take causal
relations as already decided, so to speak, and we reason on top of
them. 

As we said in Section~\ref{sec:ConcSysts}, a concurrent system is
first specified using a formal syntax, then an operational model is
built so that to describe all its possible behaviours,
and finally formal analysis is conducted on top of the model in order
to prove its properties. So, even if event structures take causal
dependencies as primitive, the difficult problem is not completely
eluded: given the system, we still have to associate it with a correct
event structure, that is an event structure whose (primitive) causal 
dependencies agree with the system behaviour.  
%
%
More precisely, we say that a model is \emph{correct} with respect to
the system if any behaviour in the model is a plausible behaviour of
the real system; we say that the model is also \emph{complete} if any
behaviour of the real system also appears in the model. 
While a model that is not correct is useless, models that are not
complete are often useful abstractions of complex systems.
Notice for instance that an empty model is correct with respect to any
system even if it is clearly useless. Any other definition stating
what kind of system information is represented in the model is
somewhat arbitrary, and it is essentially justified by a proof that 
the given model is correct (i.e. any behaviour in the model is a
plausible behaviour of the real system) and an illustration of the
reasoning and predictions about the system allowed by that model.

How to (automatically) associate an operational model to a
given system is a lively research topic in computer science. While
operational semantics based on sequential or interleaving models are
pretty standard, the case of true-concurrent/causal models is more
controversial. 
As an example, consider a software system written in the Java
programming language. A naive approach would define a programming
instruction A as dependent on an instruction B whenever B appears in
the code before A. However, such a definition would lead to an
incorrect model when B is the instruction ``run this instruction in
parallel with the code appearing in the following''. Moreover, the
sequence of program instructions written by the programmer do not in
general correspond to the sequence of instructions actually executed
by the hardware: indeed the runtime program execution is often defined
so that to allow the reordering of instructions to get optimized
execution performances. Rather than precisely modelling the reordered
runtime program, the Java programming language relies on the so-called
Java Memory Model \citep{JMM}, which defines a partial ordering called
\emph{happens-before} on all actions within a program. 
In absence of a happens-before ordering between two operations, the 
runtime execution is free to reorder them as it pleases. 
Then in defining the causal model of a given Java program, the
definition of the causal relation linking the program instructions
will then agree with (but non necessarily be the same as) the
happens-before relation defined in the specification of the Java
programming language. Moreover, the advent of new parallel hardware
(e.g. multicore CPUs or GPUs) have renewed the debate about the
identification of the key dependencies between software 
instructions, leading to different memory models for new programming
languages.

So to summarize, the difficult problem of deciding whether two
events are causally dependent or not is confined into the problem of
correctly associating an event structure to a concurrent system. 
The definition of a correct and useful model for a given concurrent
system is a lively research topic in the area of formal semantics. 
Anyway there is no causal discovery to do; the debate
generally amounts to the (somewhat arbitrary) definition of a
``precedence relation'' between system actions.

\paragraph{Causality vs Dependency.}
The examples above show that the causal relation $\leq$ of event
structures encodes any form of dependence, such as temporal and 
or spatial precedence, rather than puzzling over the causal
\emph{nature} of the link between two events. Reasoning in this way
has two benefits: 

\begin{itemize}
\item [(i)] Having a model that talks about causality/dependencies
  allows formal reasoning about such dependencies. For instance, we can
  define specification logics that help with proving system properties
  like 
  ``it is true that at any time an action A depends on an action B and
  is concurrent with an action C''. In the phone call and text
  message example above, it is true that Bob's action B2
  depends on Alice's action A1 and is concurrent with Alice's action A2, 
  whereas it is not true that Alice's action A2 depends on the action
  B1 performed by Bob before reading the text sent by
  Alice.
  So more generally, we can say that formal reasoning about
  dependencies help prove some important properties of the system (or
  of the model).  

  As another example of the impact of these models in formal reasoning, consider the reasoning of ``reductio ad absurdum'', commonly used in maths. Such reasoning entails
  \emph{indirect} proofs as ``reductio ad absurdum'' arguments, that
  derive a contradiction from a to-be-refuted false assumption.
  This proof technique also applies to the verification of concurrent
  systems, both in proving the properties of a given model and in
  proving the correctness of a general analysis technique.
  The effectiveness of indirect proofs comes form the
  fact that in the true-concurrent models described so far, given any
  pair of events $A,B$, the model precisely states 
  whether they are causally dependent, or in conflict or concurrent. 
  The key point is that there is no room for an
  unknown/yet-to-be-discovered link between them. Such a full knowledge
  is essential when conducting reductio ad absurdum arguments, since
  starting form a negative assumption such as ``$A$ and $B$ are not
  causally linked'' we can still proceed in the argument by
  distinguishing two subcases: (1) the positive assumption that $A$ and
  $B$ are in conflict and (2) the positive assumption that $A$ and $B$
  are concurrent.

\item [(ii)] The choice of encoding any dependency with the same relation
  rather than distinguishing between the natures of the dependency is
  particularly well suited to study independence of actions, which is
  at the core of concurrent systems. 
  One of the major impacts of studying independency of actions is the
  ability to optimize the execution of a software system on top of new
  multiprocessor/multicore architectures. A single processor, or a
  processor with a single core, can only execute a single instruction
  per cycle, hence if A and B are two concurrent/independent actions,
  the processor can only perform an interleaving of A and B, that is
  either the sequence A and then B or the opposite sequence B and then
  A. On the other hand, if the processor is dual-core, A and B can be
  executed simultaneously. By correctly identifying independent
  actions, we can for instance compute the ``degree of concurrency''
  of an application, and then estimate the maximal number of
  processors/core that application makes use of.  
 
\end{itemize}

It is important to observe that independence is not just the
  negation of causality. Indeed, the event structure model makes
  clear that two events that are not in the causal relation might 
  be either independent/concurrent -i.e.~they can occur in any order
  and also simultaneously---or in conflict---i.e.~they can be both
  enabled but the occurrence of one disallows the occurrence of the
  other. In other terms, independence is not just
  non-(causal)dependence, but it is both non-(causal) dependence and
  non-conflict. We shall discuss causality and independence in more
  detail later.

\paragraph{Disjunctive causality.}
We have mentioned that in prime event structures events with disjoint
causes get duplicated: for instance in the Computer Network example
above the presence of the message in Moscow is represented by two 
different events, one depending on the transit of the message through
London, and the other one depending on the transit through Paris.
As a general example, consider: 
``If Argentina or Brazil win the world cup, I'll eat my hat''. 
This case is modelled by means of two separate events: 
\begin{itemize}
\item [(i)] ``I eat my hat because Argentina won'' and 
\item [(ii)] ``I eat my hat because Brazil won''.
\end{itemize}

Notice that the two alternative causes, ``Argentina wins'' and
``Brazil wins'' are mutually exclusive.
Furthermore, any event depending on ``me eating my hat''
must also be duplicated, so that a copy of it depends on
``me eating my hat because Argentina won'' and a different copy
depends on ``me eating my hat because Brazil won''. 

The duplication of events corresponding to the 
same action with different causal histories guarantees an important
property of prime event structures: in these models the causal
history of any event is always fully and non-ambiguously known.
However, such a property implies that these models cannot represent 
the so-called inclusive disjoint causality. 
Suppose at the beginning of the European Cup I say:
\begin{itemize}
\item []  ``If Germany or Italy do not reach the final, you'll eat
your hat.''
\end{itemize}

\noindent 
If both teams don't reach the final, which one, out of the two, is the cause of your eating
your hat? There are two possible causes (that are not mutually
exclusive), that is ``Germany does not reach the final'' and ``Italy does
not reach the final''. However, from the perspective of concurrent systems, there is no real gain or interest in
establishing which one, out of the two, is the cause. 
Notice indeed that in computer science
it is often preferable to promptly rely on incomplete information
rather than to engage space/time resources in retrieving complete info.
It  sometimes happens that even if we could in principle establish
whether there is a dependence between two actions, we decide that it is not
worth looking at it in more detail.

When we are just interested in sets of \emph{possible} causes we
cannot rely on prime event structures anymore. However, 
generalized event structures has been studied by relaxing the
constraint of fully knowing the causal history of events and
by introducing some sort of incomplete information \citep{Winskel87}. 
Observe, in particular, that losing a piece of information about
dependencies is a price that we can pay when we are just 
interested on the concurrency properties of the system, that is on its
independent actions.

\medskip

\noindent Summing up, encoding any form of precedence with the same relation 
yields already a very expressive model to reason about. Moreover, 
this is good enough, as long as all these kinds of dependency do
not diverge/oppose, which appears to be the more frequent case. 
Also, extensions of event structures dealing with more
than a single ``causal relation'' are also the object of current
research. At this point, it is worth comparing this meaning of ``causality'' with the discussions happening in the philosophy of causality. 

\subsection{A comparison with concepts from the philosophy of causality}

There are a number of ways in which the meaning of causality as just discussed differ from the meaning discussed in the philosophy of causality.

\paragraph{Production and mechanisms.} In concurrent systems, while we
are interested in whether events A and B stand in a relation of
dependence, we are not interested in  other possible meanings of
causality, most notably, \emph{production}. One problem discussed in the philosophy of causality is precisely about the nature of the connection from the cause to the effect. The Salmon-Dowe mechanical model  for instance, identifies production in the exchange of conserved quantities in physical processes (see \citet{salmon84,Dow00b}). But this account has been subsequently criticised by \citet{MDC} because it is not well-suited to biology, where the production is given by bio-chemical interactions that happen between entities in complex systems. In social contexts, production is cashed out terms of interactions between individuals, or communication, role of norms and values \citep{hedstr-ylik-2010}. 

We are not saying that events in a model cannot stay in a `productive' relation, but that this is not what we are interested in establishing. We are not interested in whether event A causes event B in the sense of \emph{producing} it, for instance in the sense that a virus produces flu. Likewise, we are not interested in reconstructing the causal \emph{mechanism} that is responsible for a phenomenon, for instance when we are interested in understanding the mechanisms of spread of an infection. 
We have seen that in constructing the causal model, it is sufficient to define a ``precedence relation'' between system actions.

\paragraph{Preemption.} Likewise, the examples about disjunctive causality just mentioned have the same structure of the well-known example of  `Billy and Suzy' in the philosophy of causality. However, from the same \emph{conceptual} structure we draw different conclusions. Let us explain.

In the literature there are many variants of the example, but roughly the story goes like this. Billy and Suzy are playing throwing rocks at a bottle. Billy throws first and shatters the bottle, Suzy throws next, the rock going through exactly where the bottle was. So, had Billy missed the bottle, Suzy would have shattered it. The example shows a difficulty for the counterfactual theory of causality to correctly identify the cause that is responsible for an event. Is Suzy or not  the \emph{actual} cause of the shattered bottle? In fact, had Billy missed, she would have made it. And yet, at the same time,  Suzy is just a preempted potential cause (for a presentation and discussion, see \citet{hall-2004,menzies-2009}.

The example  certainly  shows that causal assessment in natural language is oft undecided or unclear and that counterfactual reasoning may fail to give a definite answer. It does not follow, though, that in some specific disciplines causal relations remain undecided. In law, for instance, we can decide about causal assessment based on tools that \emph{complement} counterfactual reasoning, including the whole corpus of jurisprudence or the distinction between `causes' and `conditions' \citep{hart-honore-1985}.

The majority of the (counter)examples in the philosophy of causality are meant to illustrate the difficulties for a specific theory or approach to unambiguously identify causal relations. In concurrent systems, while we can find examples sharing the same structure of Billy and Suzy, they do not illustrate the same shortcomings. The reason is that, in this context, programmers are not so much interested in deciding what causes what, but what relations holds and do not hold in order to run a programme without bugs. Admittedly, this might show that the use the term `causality' in concurrent systems is a bit overloaded with meaning.

\paragraph{Independence.}
In statistical modelling and probabilistic causality, the meaning of  independence is different from the one given earlier for concurrent systems. In these contexts independence specifically means  \emph{statistical} or \emph{probabilistic} conditional or unconditional independence between variables of interest \citep{Sup70}. In the approach known as `Granger causality', independence is used  to analyse the present state of a given variable---e.g., wealth conditions in a population of retired people---given the history of other, possibly related variables---e.g., their health conditions and the history of their wealth \citep{granger-1969}. In this way Granger tested independence: he was interested in whether information about past history of some variables is (or isn't) informative in order to determine the value of the outcome variable. Here, causality is defined through its negative: we say that A does not ``Granger-cause'' B if the history of A is irrelevant for B. 

This use and meaning of independence is clearly at variance with the use and meaning of independence in concurrent systems, where it has  to do with the (potentially) simultaneous execution of events in the model, which leads to some consequences 
in the evolution of the system, to be studied and analysed.

\paragraph{Causation by omission.}
Causation by omission happens when an absence or a non-entity supposedly cause something. For instance, not watering the plant caused it to die. Philosophers worry about cases like this because it doesn't exist a `metaphysical' causal link between the cause and the effect. One reason to insist on such a link is that often we are interested in the `productive' relation between the cause and the effect. But how can a non-event such as \emph{non} watering the plant cause its death? Possible solutions to this puzzle come for instance from legal theory, where a causal nexus between an omission and its effect can still be established, but on other grounds (e.g., legal or moral responsibility) \citep{pundik-2007}. Or from biology, where scientists notice that the omission of watering the plant causes its death because other mechanisms activate, for instance dehydration \citep{mach-04}.

\bigskip

\noindent In sum, all these considerations mark an important difference between concurrent systems and the philosophy of causality, where scholars by and large agree that causality, in various scientific contexts, involves a `dependence' component and a `productive' component \citep{hall-2004,russo-williamson-causality-07, Illari-2011, Inf-trans}. But all this is just to say that concurrent systems seem to have different worry, in spite of a similarities of types of problems.  

Of course this opens the debate whether `causality' be used improperly in concurrent systems. This may well be. The analysis above suggests that concurrent systems may as well dispose of the term `causality' and employ `dependence' instead, without loss of content in their modelling practices. Yet, to argue for a reform of technical vocabulary in concurrent systems goes well beyond the goals of this paper. 

At the same time, the present discussion paves the way for a more thorough examination of causality vs dependence, not only from conceptual point of view, but also for applications. For instance, when the formalism of concurrent systems is used to model biological systems, e.g. in system biology, the absence of a clear distinction between causality, generic dependency, precedence, necessary condition, leads to problems, since in biological systems
the notion of causality certainly does not reduce to necessary conditions/precedence relation.

We will consider already an achievement to have clarified what `causality' means and how `causal reasoning' works in this context. It will be a useful bridge between the philosophy of causality and scientific context that attracted comparatively less attention from the community.

\section{Counterfactual Reasoning in Concurrent Systems}
\label{analysis}

We now turn to the discussion about the form of causal reasoning
in concurrent systems. 
Recall, a concurrent system may evolve in many 
different ways, and its semantic model precisely describes
the (possibly infinite) set of its possible executions.
Analyzing a system amounts to study the system properties,
where a property is a proposition which holds true in any 
execution of the system.
However,  reasoning about the models of real (rather than `toy') 
concurrent systems is
in general very difficult since the size of these models is huge and
possibly unbounded. In these cases an exhaustive look at the model in
unfeasible. Available approaches include advanced model checking
techniques, that rely on abstractions based on behavioural
equivalences or suitable temporal logics \citep{ModelCheck}.
These technique can deal  with the complexity
of concurrent systems verification.
More generally, a
 rich toolbox of formal methods have been developed to automatically
reason about causal models, that is to specify and verify 
(causal) properties of (models of) concurrent systems.
These analysis techniques can be partitioned into three main
categories: the analysis conducted on the system model before the
system execution (static analysis), the analysis conducted during the
system execution (dynamic analysis, execution profiling) and the
analysis of the actual behaviour executed in a system run (trace
analysis, fault diagnosis).

In this section we discuss how the third class of techniques,
i.e., fault diagnosis or execution trace analysis, shares connections with the philosophical debate,
namely conditionals and counterfactuals.

We said earlier that models of real concurrent systems are huge and
possibly unbounded, essentially because of the nondeterminism inherent
to concurrent systems. Remind that nondeterministic executions mean
that if we run the same program several times, its execution can be
different each time. For instance, in the railway example, on one
occasion train 1 arrives before train 2, but on a subsequent occasion
train 2 arrives before train 1, and the system model meant
to describe all the possible executions. 

In order to deal with such a complexity, only parts of the model are
built, or the modelled behaviour is just an abstraction/approximation
of the actual system execution. In particular, correct system
behaviours are precisely described by the model, while system failures
are less detailed in the model. Then, if during an execution something
goes wrong, 
we need to run an \emph{ex post} analysis to
find the error in the programme. In particular we need to perform a
fault diagnosis, that is to identify the exact system behaviour and
the ``cause'' of the error. The causal model of the system then turns
out to be very useful to focus on the specific incorrect behaviour and
to reason on the chain of events that led to the error.
Interestingly, such a process involves \emph{counterfactual}
reasoning.  

We will make use of  the approach to counterfactuals developed by Nicholas \citet{rescher-2007}. We first briefly present the theory and then illustrate how the theory can be applied in the \emph{validation} and \emph{refutation} of counterfactuals. We end the section with some general conclusions about the use of counterfactuals in concurrent systems.

\subsection{Rescher's theory of counterfactuals}

A counterfactual is a subjunctive conditional where the antecedent is known or supposed to be false. Everyday language provides plenty of examples: ``if I hadn't missed the buss, I would have been on time for class''; ``had I watered the plant, it wouldn't have died'', etc. Counterfactuals are often used to reason about causes and effects, in particular about what would have happened had the putative cause not occurred. The goal of a counterfactual is then to pick out the `right' cause and we'll know that it did in case it holds true.

The problem, as is well known, is that it is easy to show that classical propositional logic is not a suitable logic for giving us truth values of counterfactuals. In fact, if we analyse subjunctive conditionals as material implications, then we are bound to consider \emph{any} counterfactual as true, given that the antecedent is false---this a consequence of the paradoxes of material implication. 

This situation led the philosopher and logician David Lewis to explore a different path: counterfactuals are regimented by a possible-worlds semantics \citep{Lew83,Lew00}. Simply put, a possible-world semantics rests on the assumption of the existence of a \emph{plurality} of worlds, among which there is also our \emph{actual} world. This position is also known as modal realism. Counterfactual evaluation is carried out by  comparing worlds with each other on the basis of their similarity or closeness to our actual world: the closer the world is, the more similar it is to the actual world. The truth of a counterfactual  is then ascertained by an `inspection' of what happens in other possible worlds, and according to the rules of counterfactual dependence as defined in the Lewis-Stalnaker semantics \citep{stalnaker-1968,stalnaker-1975,lewis-1973}. `Lewisian' counterfactuals gave rise to an incredibly vast literature, defending, refining, and criticising the account, to the point that counterfactual reasoning is usually associated with Lewis' possible worlds approach, while alternative theories are often neglected. One such theory is Nicholas Rescher's, that we present next.

Rescher criticises Lewis' theory because of its (unnecessary) metaphysical baggage and also because of its \emph{practical} difficulties, for instance how to get from one world to another or how to implement proximity between worlds. Rescher's account, on the contrary, is capable of making sense of counterfactual validation in a way that is logically precise and rigorous, and that is metaphysically  parsimonious.  In fact, Rescher claims his theory to be \emph{epistemic} rather than metaphysical: while Lewis-Stalnaker invoke possible worlds and possible objects, Rescher invokes sets of \emph{beliefs}, ``which can straightforwardly, after all, be finite in scope, and indeed sometimes even inconsistent'' \citep[p.166]{rescher-2007}.

To begin with, Rescher adopts a slightly different characterisation of `counterfactual': a counterfactual ``purport[s] to elicit a consequence from an antecedent that is a belief-contradicting supposition, on evidence that it conflicts with the totality of what we take ourselves to know'' \citep[p.74]{rescher-2007}. This definition avoids reference to events and to the truth or falsity of the antecedent. All is phrased in epistemic terms, appealing to beliefs and evidence. 

The formalism required by Rescher is not very heavy. He represents a counterfactual as $p \lbrace B \rbrace \rightarrow q$, which is read ``If $p$ were true, which we take not to be so---not-$p$ being a member of the set of our pertinent beliefs $B$ (so that $\neg p \in B$)---then $q$ would be true'' \citep[p.81]{rescher-2007}. The idea is that a counterfactual holds, or is validated, when the belief-contradicting supposition which is at stake will yield its consequent as a \emph{deductively valid} conclusion. To perform the demonstration, we will have to add suitable supposition-compatible beliefs.  The validation of a counterfactual has therefore the following generic structure: (disbelieved antecedent $+$ certain accepted beliefs) $\vdash$ consequent. Rescher calls it a \emph{derivability construal of counterfactuals}. 

The consequent $q$ is derivable from the combination of $q$ plus some appropriately subsection of the set of background beliefs. The problem, of course, is to determine which beliefs to choose. These background beliefs are called  `belief-compatible assumptions'. The choice of these assumptions is guided by a principle of conservation of information, namely ``prioritising  our beliefs in  point of generality of scope and fundamentality of bearing'' \citep[p.105]{rescher-2007}.

Counterfactual validation is then about restoring consistency in an optimal way by prioritising information. For instance, conceptual relations have priority over mere facts, norms of practice will advantage some facts over mere matters of brute contingency. This idea is made more precise and sophisticated by spelling out the acronym `MELF'. MELF stands for \textbf{M}eaning, \textbf{E}xistence, \textbf{L}awfulness, \textbf{F}act. It indicates how precedence and prioritisation work in the absence of case specific specifications to the contrary.
The derivability of $q$ depending on B (i.e.~background context) is done using MELF, which guides us in choosing what background beliefs to use and what beliefs to discard in the derivation.

\noindent
Consider the following example borrowed from
\citet[p.106]{rescher-2007}:
\begin{center}
\begin{minipage}{12cm}
\begin{center}
``If this rubber band were made of copper, it would conduct electricity''.
\end{center}
\end{minipage}
\end{center}
This counterfactual arises in an epistemic context where the following
beliefs are salient:
\begin{enumerate}
\item This band is made of rubber (factual belief).
\item This band is not made of copper (factual belief).
\item This band does not conduct electricity (factual belief).
\item Things made of rubber do not conduct electricity (lawful belief).
\item Things made of copper do conduct electricity (lawful belief).
\end{enumerate}
We now negate the second belief: this band is made of copper. 
To restore consistency between the initial set of beliefs, 
we'd have to choose whether to keep 3.~or 5., that is between a
particular feature of this band and a general fact about copper
things. Since 5. is systematically more informative than 3., it has
higher priority, hence we keep 5.~in the background beliefs and we 
therefore accept the counterfactual.

%

An important remark in using MELF is the following. While in \emph{factual} contexts we give priority to \emph{evidence}---i.e.~the most supported proposition is the one that is most strongly evidenced---in \emph{counter}factual contexts, we give priority to `fundamentality' (rather than evidentiation)---i.e.~aspects about meaning, existence, and lawfulness. 

For the interested reader, \citet[ch.11]{rescher-2007} book provides  more technical details on the logics to be used for MELF considerations in derivability; in particular, Rescher discusses the inductive character of his derivability construal, and also other important properties such as monotonicity and transitivity (it is non-monotonic and transitivity fails).

\bigskip

\noindent With this theoretical background, we can now illustrate counterfactual reasoning at work in a concurrent systems example.


\subsection{Application of counterfactual reasoning in concurrent systems}

We illustrate counterfactual reasoning taking inspiration from a real case, Rover on Mars. A precise account of what happened to Mars Pathfinder is out of our scope.  We just describe here an example containing an erroneous behaviour  with the aim of illustrating counterfactual reasoning over causal  models. The real case is documented
in \citet{MarsRover}
These are the salient facts:

\paragraph{Rover on Mars.}
On the 4th July 1997 the Mars Pathfinder bounced onto the Martian
surface  surrounded by airbags. After landing, it deployed the
Sojourner rover which started gathering and transmitting voluminous
data back to the Earth. 
But a few days into the mission, the spacecraft began
 experiencing system resets. After the failure, NASA engineers spent
 hours and hours running the system on the exact spacecraft replica in
 their laboratory, attempting to replicate the precise
 conditions under which they believed that the reset occurred. 
 When they finally reproduced a system reset on the replica, the
 analysis of the computation trace revealed a well known concurrency
 bug, a so-called priority inversion problem. 

The essence of the Mars Pathfinder experience can be illustrated in
terms of counterfactual reasoning on top of
the concurrency models discussed in previous sections. 
The software system controlling the Sojourner rover is a
concurrent system that carries out and coordinates a number of
parallel activities. For the sake of simplicity we can assume that
the following event structure is a (minimal) portion of the whole
operational model describing the rover's behaviour:
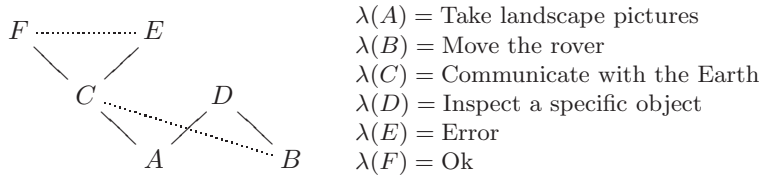
\begin{center}
\vspace*{-1.6cm}
\begin{tabular}{cc}
\mbox{
\xymatrix@C=4mm@R=4mm{
 F \ar@{.}[rr]&   & E &   & \\
   & C\ar@{-}[ul]\ar@{-}[ur] &   & D & \\
   &   & A\ar@{-}[ul]\ar@{-}[ur] &   & B\ar@{-}[ul]\ar@{.}[ulll]
}}
&
{\small
$
\begin{array}{l}
\\[1.3cm]
\lambda(A)=\mbox{Take landscape pictures} \\
\lambda(B)=\mbox{Move the rover}\\
\lambda(C)=\mbox{Communicate with the Earth} \\
\lambda(D)=\mbox{Inspect a specific object}\\
\lambda(E)=\mbox{Error} \\
\lambda(F)=\mbox{Ok}
\end{array}
$
}
\end{tabular}
\end{center}
The model shows that initially the rover can move (event $B$) and at
the same time it can take pictures (event $A$). We can think the
event $D$ as the rover inspecting a specific object that had been
previously identified in a picture and after the rover have moved next
to it, i.e.  $D$ ``causally depends'' on both $A$ and $B$. Moreover,
the model specifies that moving the rover (event $B$) is in conflict
with a communication with the Earth (event $C$), which means that the
execution of $B$ prevents the execution of $C$, and viceversa.
Therefore, even if initial independent actions $A$ and $B$
could be in principle executed in any order or even concurrently,
different choices in their execution order 
have different impacts on the possibility of executing $C$.
Finally, the model shows that after a communication with the Earth,
the system enters either a correct state $F$ or an error state
$E$. Let suppose that such a choice depends on the presence of light
on the Martian surface: if after a communication with the Earth the
rover is in the darkness, then it enters an error state.

Let assume that this event structure is a correct model for the
Sojourner rover, that is all the behaviours in the event structure are
plausible rover's behaviours. Than the correctness of the \emph{model}
shows that the \emph{system} itself is not correct: indeed, a system
is said to be 
correct whenever no system run can end up in an error state, which is
clearly not the case of the rover. Notice that such a situation is not
so odd: the entire model of the rover system has millions of states,
hence it is usually unfeasible to generate (and inspect) the whole
model. Instead, only parts of the model are built up, or it is
generated only an approximation of the concrete model, taking into
account the risk of leaving unnoticed an error state, typically hidden
in an infrequent system behaviour.
That's exactly what happened to the Mars Pathfinder: the engineers did
not realize the system error and sent the rover to Mars.

Once activated on the Martian surface, the system started to run
according to its model. There were three initial possibilities:
executing $A$, or executing $B$, or executing both $A$ and $B$
concurrently. The rover on Mars executed $A$. After that, the rover
could choose to execute either $B$ or $C$; it executed $C$ and then
moved to the error state $E$ since the rover found itself into the
darkness. 

\bigskip

\noindent We now distinguish two types of use of counterfactuals: validation and refutation.

\paragraph{Counterfactual validation.} The precise sequence of actions 
performed by the rover was not known by the NASA engineers on the
Earth. They only new that the rover had started its execution
according to the installed software, meaning that it  was
executing one of the possible behaviours they had programmed.
It has been reported that after the unexpected failure, the engineers
``spent hours and hours running the system on the exact spacecraft
replica in their laboratory, attempting to replicate the precise
conditions under which they believed that the reset occurred''. 
This can be rephrased in our example as the engineers setting out all
the possible behaviours in the system model until they found a
behaviour ending up in the error state reported by the rover.
Their explanation of the problem could then be stated as the following
counterfactual: 
\begin{center}
\begin{minipage}{9cm}
\begin{center}
``Since it was dark, if the first rover's action had been $B$,\\
 it would not have entered the error state.''
\end{center}
\end{minipage}
\end{center}
Now, using Rescher's MELF theory, our beliefs are as follows:
\begin{enumerate}
\item It was dark.
\item The rover did not perform $B$ as its first action.
\item The rover performed $C$.
\item The rover ended up in $E$.
\item The execution of $B$ prevents the execution of $C$.
\item If $E$ is executed, then it is dark and $C$ has been previously
  executed.
\end{enumerate}
where 1-4 are ``Facts'', while 5-6 are ``Laws''.
In order to validate the counterfactual let assume not-2, 
then 3 and 5 become incompatible. MELF theory prioritizes  the lawful
5 over the merely factual 3. We then retain 5, which is however
incompatible with 4+6 Again we reject the weaker factual 
4 obtaining the desired conclusion. Notice that we could have used 
a slightly different belief: (5bis) The execution of $B$ prevents the
execution of $E$. This choice leads to a simpler counterfactual
validation: assume not-2, then 5bis and 4 are incompatible,
and 4 must be rejected since it has lower precedence.

\medskip
As a further example, given the actual rover execution, consider the
following counterfactual:
\begin{center}
\begin{minipage}{8cm}
\begin{center}
``If there had been light, the rover would not have entered the error state.''
\end{center}
\end{minipage}
\end{center}
Notice that the ``cause'' of the rover error is not (just) the
darkness, but (also) the fact that, after $A$, in the choice between
executing $B$ or $C$, the rover executed $C$. In other terms, even if
there had not been light (which was indeed the case), the rover could
not have entered the error state.
Anyway, this counterfactual is valid. Indeed, consider the following
beliefs: 
\begin{enumerate}
\item It was dark.
\item The rover ended up in $E$.
\item If $S$ is executed, then it is dark.
\end{enumerate}
Assuming not-1 we have that 2 and 3 become incompatible, and the
factual 2 must be rejected in favour of the lawful 3 The dismissal
of 2 proves the counterfactual. 

The puzzling thing behind this counterfactual is that the cause of the
event $E$ is twofold (i.e., the darkness \emph{and} the occurrence of
$C$ instead of $B$) while the assumption of the counterfactual just
falsifies a single cause. The counterfactual validity then comes since
a conjunction can be negated by just negating one of its
operands. The dual case, that is disjunction, is more delicate, and
has rather to do with counterfactual refutation, which we discuss next.

\paragraph{Counterfactual refutation.}
Let now assume that the actual rover execution on the Martian surface
had been the following, which is still consistent with the model
above: first $B$, then $A$ and finally $D$, everything in the
darkness. Let consider the following counterfactual:
\begin{center}
\begin{minipage}{8cm}
\begin{center}
``If the first rover action had been $A$, it would have ended up in
the error state.''
\end{center}
\end{minipage}
\end{center}
We know that this counterfactual is false since, even if the sequence starts with $A$, the Rover still has to choose between $B$ and $C$, and only choosing $C$  would lead to an error.

In order to show how to precisely refute this counterfactual according
to Resher's theory, we first discuss the following example, taken form
\citet[p.123]{rescher-2007}: consider the following two counterfactuals
\begin{description}
\item[(C1)] If Napoleon were Julius Caesar, then Napoleon would be
  dead by 100 CE (since Caesar was).
\item[(C2)] If Napoleon were Julius Caesar, then Caesar would be alive
  in 1800 CE (since Napoleon was).
\end{description}
These two conditionals seem innocuous when taken separately, but they
are not cotenable since they constitute an inconsistently conflicting
pair. Indeed, using MELF approach, they cannot be validated: consider
the following beliefs
\begin{enumerate}
\item Caesar was dead by 100 CE.
\item Napoleon was alive in 1800 CE.
\item If X=Y, then whatever holds of X will hold of Y.
\item Napoleon is not Caesar
\end{enumerate}
Let assume not-4. Then B1=\{not-4, 3, 2\} is a consistent subset of
beliefs which gives ``Caesar was alive in 1800 CE'', while
B2=\{not-4, 3, 1\} is a different subset of beliefs which gives
``Napoleon was dead by 100 BC''. Rescher's MELF theory requires the
consequence to be obtained from \emph{every} consistent subsets of
beliefs. So (C1) cannot be validated since ``Napoleon was dead in 100
CE'' is only obtained from B2 whereas form B1 (actually from 2) we
have that ``Napoleon was not dead in 100 CE'' (since he was alive in
1800 CE). Similarly (C2) can be obtained from B1 while B2 proves its
denial. Notice that the only counterfactual we can validate is
\begin{description}
\item[(C3)] If Napoleon were Julius Caesar, then either Napoleon would
  be dead by 100 CE or Caesar would be alive in 1800 CE.
\end{description}
Coming back to the rover example, the valid couterfactual is 
\begin{center}
\begin{minipage}{8cm}
\begin{center}
``If the first rover action had been $A$, it would have end up either
in error or in the correct state $D$''.
\end{center}
\end{minipage}
\end{center}
More precisely, for any set of beliefs, we have a (salient) law
stating that after $A$ the rover can execute either $C$ (leading to
the error) \emph{or} $B$ (preventing the error). The presence of such
a disjunction actually prevents the validation of the original
counterfactual. Notice also that this very same argument prevents
the validation of its \emph{denial}, that is ``If the first rover
action had been $A$, it would \emph{not} have end up in the error
state.''
Anyway, how do we prove that a counterfactual is false in MELF theory?
Recsher only seems to refer to proving the counterfactual negation or
denial, but we have seen that it might not always be possible.
On the other hand, since a counterfactual C must be verified in
\emph{every} consistent subset of (salient) beliefs, in order to
reject it it is sufficient to show a consistent subset of beliefs that
implies the denial of C. Notice that it is still different from
validating the denial, which would indeed require a proof of the
denial for every subsets of beliefs rather than for just one.

In other words, proving a counterfactual amounts to deal with 
a universal quantifier (i.e., validate it for every consistent subset
of beliefs), while to refute a counterfactual we only have to 
deal with an existential quantifier (i.e., show a consistent subset of
beliefs that implies the denial).  
For instance, in the Napoleon-Caesar example above, the subset B1,
respectively B2, can be used to prove that the counterfactual (C1),
respectively (C2), is false.
Interestingly, in concurrent systems, operational models offer 
a very effective way to prove that a counterfactual is false: indeed
it is sufficient to show a specific behaviour that is allowed by the
model and where the counterfactual is false. The behaviour $A$, then
$B$ then $D$ is indeed a plausible behaviour that refutes the
counterfactual, while the behaviour $A$ then $C$ then $E$ is a
plausible behaviour that refutes the counterfactual's denial.

%

\subsection{Philosophical morals}

Before closing this section, we draw some general  morals about the use of counterfactual reasoning in concurrent systems using Rescher's approach.

Rescher's account is versatile and well fits the needs of counterfactual
reasoning in concurrent systems. More precisely, in Rescher's
theory we can distinguish between different types of belief,
on the other hand using operational models we can rely on
simplified distinction just between `facts' and `laws'.
In particular, `facts' correspond to assertions such as ``event E
occurred'' or ``event E  did not occurred'', and `laws'
correspond to the relations  between events as they are described in
the model,  e.g.~``A and B are independent / dependent / in conflict''
(these relations can be   written as implications between occurrences
of events). 
Such a clear distinction is important because we have that using these
models in the derivation of the proof of a counterfactual 
we can always decide the priority of beliefs. This is clearly at
variance with Lewis' theory, where the similarity between possible
worlds is taken as primitive and the criteria to order worlds
according to their similarity remains an open problem in the
approach. 

Being able to use `laws' in the proofs is an advantage also for the
following reason. Given the nature of the model, for each couple of
events, it is  known what relation they stand in, as the programmer
decides about such relations. Consequently, all the `laws'
connecting all the couple of events are known from the
model. Counterfactuals can then be validated by combining salient laws
into  a well-constructed proof. Conversely, given that the model
describes all the  possible system executions, a counterfactual can be
rejected by showing a case, namely one possible execution that
violates the  counterfactual.

\section{Conclusion}

We presented an account of causality in an important field of computer science,
that is the modelling of concurrent systems. These systems,
either software, hardware or even biological systems, can be thought of 
as sets of activities that run in parallel with possible occasional 
interactions. 

The formalization of concurrent systems is an interesting area to investigate the meaning and use of causal concepts. One reason is that these systems have important practical applications since concurrent software systems and parallel hardware are pervasive. Another reason is that the literature in computer science customarily use causal terms, but a systematic investigation has not been carried out so far.

We first provided a crash course in concurrent systems to set the ground for a discussion of the meaning and use of causal concepts therein. One result is that the use of the term `cause' or `causality' in this area is perhaps not fully justified, as it could be replaced with `dependence' or 'precedence' with no loss of content or informativity. In fact, `causality' or `dependence' denote an ordering between event occurrences. But there is nothing, in such a characterization, which is specifically \emph{causal}, in the sense that causes \emph{produce} their effects.

An important feature of causal reasoning in concurrent systems is the use of counterfactuals. We examine how counterfactuals are used in concurrent systems and we illustrate applications using Nicholas Rescher's theory. This also marks a difference with the philosophical literature, that mainly focused on Lewis' account better-known account based on possible-world semantics.
   
So there are a number of ways in which the use of causal terms in concurrent systems distances itself from the traditional `hot' topics in the philosophy of causality: production, mechanisms, causation by omission. Yet, our goal in the paper is not to call for a terminological change in the field of concurrent systems. We think that at this stage a rounded discussion about similarities and dissimilarities with parallel debates happening in the philosophy of causality \emph{is} already a contribution.

\bigskip

{\small \noindent \textbf{Acknowledgements} F.~Russo is currently Marie Curie Pegasus Fellow, for which financial support from the FWO-Flanders is gratefully acknowledged.}
\bibliography{refs}

\end{document}